\documentclass[aps,prl,twocolumn,superscriptaddress,amsmath,showpacs,amssymb]{revtex4}
\usepackage{epsfig}

\newcommand{\Teff}{T_{\rm eff}} 
\newcommand{\Xin}{X^\infty} 
\newcommand{\tw}{\ensuremath{t_\mathrm{w}}}

\newcommand{\upd}{{\rm d}}

\newcommand{\Ct}{\widetilde{C}}
\newcommand{\chit}{\widetilde{\chi}}
\renewcommand{\r}{r}
\newcommand{\Y}{Y}

\newcommand{\ie}{{\em i.e.}}

\begin{document} 
 
\title{Reply to Comment on ``Fluctuation-dissipation relations in the 
non-equilibrium critical dynamics of Ising models''} 
 
\author{Peter Mayer}
\affiliation{Department of Mathematics, King's College, Strand,
London, WC2R 2LS, UK}

\author{Ludovic Berthier}
\affiliation{Theoretical Physics, University of Oxford, 1 Keble Road,
Oxford, OX1 3NP, UK}
\affiliation{Laboratoire des Verres, Universit\'e Montpellier II,
34095 Montpellier, France}

\author{Juan P. Garrahan}
\affiliation{Theoretical Physics, University of Oxford, 1 Keble Road,
Oxford, OX1 3NP, UK}
\affiliation{School of Physics and Astronomy, University of Nottingham,
Nottingham, NG7  2RD, UK}

\author{Peter Sollich}
\affiliation{Department of Mathematics, King's College, Strand,
London, WC2R 2LS, UK}

\date{\today} 
 
\begin{abstract} 
We have recently shown that in non-equilibrium spin systems at
criticality the limit $\Xin$ of the fluctuation-dissipation
ratio $X(t,\tw)$ for $t\gg\tw\gg 1$ can be measured using
observables such as magnetization or energy [Phys.\
Rev.\ E {\bf 68}, 016116 (2003)]. Pleimling argues in a Comment
[cond-mat/0309652] on our paper that for such observables correlation
and response functions are dominated by one-time quantities dependent
only on $t$, and are therefore not suitable for a determination of
$\Xin$.  
Using standard scaling forms of correlation and
response functions, as used by Pleimling, we show that our data do
have a genuine two-time dependence and allow $X(t,\tw)$ and $\Xin$
to be measured, so that Pleimling's criticisms are easily refuted.
We also compare with predictions 
from renormalization-group calculations, which are consistent with our
numerical observation of a fluctuation-dissipation plot for the
magnetization that is very close to a straight line. A key point
remains that coherent observables make measurements of $\Xin$
easier than the traditionally used incoherent ones, producing
fluctuation-dissipation plots whose slope is close to $\Xin$ over
a much larger range.
\end{abstract} 
 
 
\maketitle 
 
In our recent paper~\cite{MayBerGarSol03} we analysed
out-of-equilibrium fluctuation-dissipation (FD) relations in
ferromagnetic spin systems quenched to criticality. One measures a
response at time $t$ to a perturbation at an earlier time $\tw$, the
waiting time, and compares with the corresponding two-time
correlation.  The FD ratio (FDR), $X(t,\tw)$, then captures how much the
response deviates from what would be expected in equilibrium. In many
situations, $T/X$ can be thought of as an effective temperature
$\Teff$ governing the out-of-equilibrium
dynamics~\cite{CugKurPel97}. In the context of critical dynamics the
role of the long-time limit value $\Xin$ of $X(t,\tw)$ for
$t\gg\tw\gg1$ as a universal amplitude ratio has been emphasized, see
e.g.~\cite{GodLuc00b}.

Exact results in~\cite{MayBerGarSol03} for the $d=1$ case with $T_c=0$, 
showed that $\Xin$ is identical
for all spin observables, whether one considers incoherent
(short-range) observables or coherent (long-range) ones such as the
total magnetization. For the coherent case, the limiting FD plot at
long times is in fact a straight line, from which $\Xin$ can be
determined trivially. Similar results were found for the case of bond
observables, where the incoherent observable corresponds to an
indicator for a local domain wall while the coherent observable is the
total energy.  $\Xin$ was again found to be identical for these
observables. The advantage of using coherent observables is much more
dramatic here: for the incoherent case the window in the FD
plot where the slope is close to $\Xin$ shrinks to zero with
increasing times, while for the coherent observable one finds again a
straight line FD plot of slope $\Xin$. Numerical simulations in
$d=2$ strongly suggested that these results carry over to higher
dimensions. In particular, FD plots for both the magnetization and the
energy were numerically indistinguishable from straight lines. We also
found the resulting values for $\Xin$ to be equal within numerical
error, suggesting that there may be a well-defined
effective temperature $\Teff$ for a broad range of observables.

In a Comment~\cite{Pleimling03}, Pleimling argues that 
magnetization and energy are unsuitable for
measuring $\Xin$ because their correlation and response functions
are dominated by one-time contributions depending only on the
measurement time $t$. We show in
this reply that Pleimling's remarks trivially apply in the regime
$t\gg\tw$, but say nothing about the regime where $t$ and $\tw$ are of
the same order. It is in this regime that our numerical data were
taken, and so they do carry non-trivial two-time information. Pleimling
also argues that our results are in ``marked contrast'' with the results of
renormalization group (RG) calculations~\cite{CalGam02}. We show
explicitly that the RG results are in excellent agreement,
predicting a limiting FD plot for the magnetization which is very
close to a straight line. We also answer an ``interesting question'' 
raised, but left unanswered, by Pleimling.

We begin by reviewing the construction of the FD plots from which we
determine $X(t,\tw)$, since Pleimling argues that the introduction 
of some one-time quantities render our plots ``unsuitable''. 
Consider a connected two-time correlation
function $C(t,\tw)=\langle A(t) B(\tw) \rangle -\langle A(t) \rangle
\langle B(\tw) \rangle$, with $A,B$ two observables, and the
conjugate response $R(t,\tw)=T \left. \delta \langle A(t) \rangle/
\delta h_B(\tw) \right|_{h_B=0}$. Here $h_B$ is the field
thermodynamically conjugated to $B$ and a factor of $T$ has been
included in the response. The non-equilibrium FDR $X(t,\tw)$ is
defined via
\begin{equation}
  R(t,\tw) = X(t,\tw) \, \frac{\partial}{\partial \tw} C(t,\tw) .
  \label{equ:FDT}
\end{equation}
This relation can be cast in terms of the step response $\chi(t,\tw) =
\int_{\tw}^t \upd t' \, R(t,t')$, \ie\ the response to a field
$h_B$ switched on at $\tw$ and kept constant since:
\begin{eqnarray}
  \frac{\partial}{\partial \tw}\chi(t,\tw) &=& - R(t,\tw) = - X(t,\tw)
 \, \frac{\partial}{\partial \tw} C(t,\tw) \nonumber \\ &=& X(t,\tw)
 \, \frac{\partial}{\partial \tw} [C(t,t)-C(t,\tw)] .
  \label{equ:FDTdC}
\end{eqnarray}
Two things are important to note. First, the correlation $C(t,\tw)$ is
a connected one, see the definition above. Second, it is physically
sensible to compare the integrated response $\chi(t,\tw) =
\int_{\tw}^t \upd t' \, R(t,t')$ to the integral $\int_{\tw}^t \upd t'
\, \frac{\partial}{\partial t'} C(t,t') = C(t,t)-C(t,\tw)
\equiv \Delta C(t,\tw)$, rather
than just to $C(t,\tw)$. These observations are irrelevant in the
usual situation of incoherent spin observables, for which one-time
correlations are constant, but are important in our case where they do
change in time (a situation sometimes referred to as physical
aging). This point is discussed in detail in
Refs.~\cite{MayBerGarSol03,FieSol02,BuhGar02,SolFieMay02}.

From~(\ref{equ:FDTdC}) it follows that a parametric plot of
$\chi(t,\tw)$ versus $\Delta C(t,\tw)$ has slope $X(t,\tw)$.  This is
obvious if $\tw$ is varied along the curve while $t$ is held
fixed. However, if a series of such plots converges to a limit plot
for $t\to\infty$, then this limit plot and its slope $X$ can clearly
be obtained by varying either $\tw$ or $t$, as long as both times are
large. For shorter times or if no limit plot exists, plots where $t$
is varied and $\tw$ is fixed will not in general have a slope related
to $X$, whether one plots $\chi(t,\tw)$ versus $C(t,\tw)$ or versus
$\Delta C(t,\tw)$. However, in the simple case
where $X(t,\tw)$ is constant, one has from~(\ref{equ:FDTdC}) that
$\chi(t,\tw) = X \Delta C(t,\tw)$, so that a $(\Delta C,\chi)$-plot
does have the correct slope. A $(C,\chi)$-plot does not, on the
other hand, since $(\partial \chi/\partial t)(t,\tw) = X (d/dt) C(t,t)
- X(\partial C/\partial t)(t,\tw) \neq -X(\partial C/\partial
t)(t,\tw)$. This lends further support to our choice of plotting
$\chi$ versus $\Delta C$ rather than $C$~\footnote{When using {\em
normalized} FD plots, the choice of $\Delta C$ vs $C$ is immaterial,
since $\Delta C(t,\tw)/C(t,t)=1-\Ct(t,\tw)$ differs from
$C(t,\tw)/C(t,t)=\Ct(t,\tw)$ only by a constant.}.

For systems where $C(t,t)$ does not converge for $t \to \infty$ it is
convenient to consider normalized functions $\chit(t,\tw)=
\chi(t,\tw)/C(t,t)$ and
$\Ct(t,\tw)=C(t,\tw)/C(t,t)$~\cite{MayBerGarSol03,FieSol02,SolFieMay02}. 
According to~(\ref{equ:FDTdC}) these are also linked by
\begin{equation}
  \frac{\partial}{\partial \tw}\chit(t,\tw) =
X(t,\tw) \, \frac{\partial}{\partial \tw} [1-\Ct(t,\tw)] .
  \label{equ:FDTdCnorm}
\end{equation}
Again, $X$ is the slope of a plot of $\chit$ vs $1-\Ct$.  If a limit plot
is approached for large times, either $t$ or $\tw$ can be varied to
obtain this plot. Explicitly, if for large times $X$ depends on $t$
and $\tw$ only through $\Ct(t,\tw)$, the shape of the limit plot
follows by integration of~(\ref{equ:FDTdCnorm}) as 
\begin{equation}
  \chit(t,\tw)=\int_{\Ct(t,\tw)}^1 \upd \Ct \, X(\Ct) .
  \label{equ:FDTint}
\end{equation}
In equilibrium $X(t,\tw) \equiv 1$ and one recovers the standard FDT
relation $\chit(t,\tw)=1-\Ct(t,\tw)$.

\begin{figure}[t]
\begin{center}
\epsfig{file=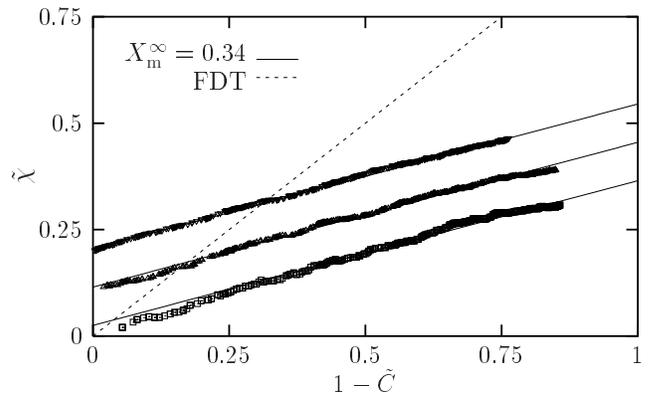,width=8.5cm}
\caption{Normalized FD plot for magnetization in the $d=2$ Ising model
at $T_c$ for times $\tw=46$, 193, and 720 (bottom to top). Curves 
have been vertically shifted by 0, 0.1, and 0.2 for clarity.
The convergence for large $\tw$ to an almost straight line of slope
$\Xin \approx 0.34$ is evident.}
\label{fig:norm_data}
\end{center}
\end{figure}

In~\cite{MayBerGarSol03}, we showed FD plots for the total
magnetization $M=\sum_i s_i$ (\ie\ $A=B=M$ above) and energy
$E=-\sum_{(ij)} s_i s_j$ for a $d=2$ system of Ising spins $s_i$
quenched to its critical temperature.  These were produced by varying
$t$ at several fixed $\tw$, and without normalization.  While a priori
the slope of the plot does then not necessarily correspond to $X$, the
numerical data for $\chi(t,\tw)$ vs $C(t,t)-C(t,\tw)$ fall on a
straight line.  Normalization only shrinks both axes of the plot in a
$t$-dependent manner.  The normalized plots will thus have the same
slope, as shown explicitly in Fig.~\ref{fig:norm_data}.  The data clearly
point towards the existence of a limit plot for large times which must
be very close to a straight line.  The asymptotic FDR $\Xin$, which
is obtained for $t\gg\tw\gg 1$, \ie\ $\Ct\to 0$, is the slope at the
end point of the limit plot (see the sketch in
Fig.~\ref{fig:sketch}).  Our data do not reach this end point, but RG
calculations (see below) show that the slope should remain constant on
approaching it.  $\Xin$ can therefore be determined from the slope
in the central part of the plot (\ie, the regime $t \gtrsim \tw \gg 1$).

We now summarize the scaling relations used by Pleimling, taking as he
did the case of the total magnetization as an example. For large
times, one expects the two-time autocorrelation of $M$ to scale as
\begin{equation}
C(t,\tw) = \tw^{a+1}f_C(t/\tw) ,
\label{equ:Cscaling}
\end{equation}
with $a$ expressed in terms of standard critical exponents as
$a+1=(2-\eta)/z=(d-2\beta/\nu)/z$. The scaling function $f_C$ decays
as $f_C(\r) \sim \r^{\theta'}$ for large $\r=t/\tw$ with
$\theta'\approx 0.19$ for the $d=2$ Ising model; in
the limit $\r\to 1$, $f_C(\r)$ has to tend to a constant to have
consistency with the scaling of the equal-time correlation $C(t,t)\sim
t^{(d-2\beta/\nu)/z}=t^{a+1}$. A similar scaling relation holds for
the response, $R(t,\tw)=\tw^a f_R(t/\tw)$.  As a result, $X(t,\tw)$
becomes for large times a function of $\r$ only~\cite{GodLuc00b}; this
is confirmed by explicit RG calculations~\cite{CalGam02}. The
normalized two-time correlation
\begin{equation}
\Ct(t,\tw) = \frac{C(t,\tw)}{C(t,t)} =
(\tw/t)^{a+1}\frac{f_C(t/\tw)}{f_C(1)} ,
\end{equation}
likewise only depends on $\r=t/\tw$ only. Eliminating $\r$, $X$ can be
expressed for large times as a function of $\Ct$. As discussed above,
it follows that a plot of $\chit$ versus $1-\Ct$ must approach a
limiting shape for large times; this is consistent with our numerical
data in Fig.~\ref{fig:norm_data}. Explicitly, if $\Ct(\r)$ and $X(\r)$
are known then the limit plot is from~(\ref{equ:FDTint})
\begin{equation}
\chit(\Ct) = \int_{1}^{\r(\Ct)} \upd\r
\left(-\frac{\upd\Ct}{\upd\r}\right) X(\r)
  \label{equ:FDTintr}
\end{equation}
where $\r(\Ct)$ is the inverse function of $\Ct(\r)$ and the minus
arises because $\Ct(\r)$ is a decreasing function.

\begin{figure}[t]
\begin{center}
\epsfig{file=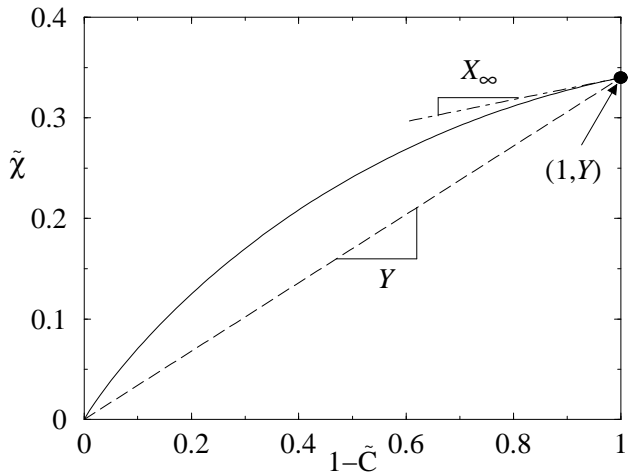,width=8cm}
\end{center}
\caption{Sketch of a limiting normalized FD plot (solid line). The
asymptotic FDR $\Xin$ is the slope of the tangent at the top right
end point of the plot (dot-dashed). This end point is at
$(1-\Ct,\chit)=(1,\Y)$, where $\Y$ is the axis-ratio of the plot or,
alternatively, the slope of the dashed line connecting the end point
to the origin.
\label{fig:sketch}
}
\end{figure}

Pleimling deduces from~(\ref{equ:Cscaling}) that, in the limit
$t\to\infty$ at {\em fixed} $\tw$, $C(t,t)-C(t,\tw)\approx C(t,t)$
because $C(t,\tw)\ll C(t,t)$. This is correct, but not surprising.  As
stated above, one generically expects that
$\Ct(t,\tw)=C(t,\tw)/C(t,t)\to 0$ for $t\gg\tw$. For the step response
$\chi(t,\tw)$ he shows similarly that this becomes independent of
$\tw$ for $t\to\infty$ and grows with the same power law as $C(t,t)$,
so that the ratio $\chit(t,\tw)=\chi(t,\tw)/C(t,t)$ approaches a
constant which we shall call $\Y$ \footnote{The limit $t\to\infty$
here has to be understood as taken {\em after} the zero-field limit
$h\to 0$.  As Pleimling discusses, if one considers instead a fixed
nonzero field then the response eventually becomes nonlinear, on a
timescale for $t$ which diverges for $h\to 0$.  While correct, this
observation is irrelevant for the present discussion: we have
checked carefully that all our data are taken in the time window where
the response is linear, as Pleimling's Fig.~2 also confirms.}. 
Summarizing, for $t\to\infty$ at fixed
$\tw$, one has $\Ct\to 0$ and $\chit\to \Y$.  Referring to
Fig.~\ref{fig:sketch}, Pleimling's statements thus fix a single point
on the limiting normalized FD plot, namely its ``end point'' on the
right.  Geometrically, $\Y$ is the axis ratio of the FD plot. It is
important to stress that Pleimling's reasoning says nothing about the
rest of the limiting FD plot, which corresponds to the time regime
where $t$ and $\tw$ are of the same order: his limit $t\to\infty$ at
fixed $\tw$ always implies the assumption $t\gg\tw$.  It is also clear
from Fig.~\ref{fig:sketch} that the axis ratio $Y$ and the asymptotic
slope $\Xin$ of the FD plot are not in general related. This
answers the ``interesting question'' which Pleimling poses at the end
of his Comment: $Y$ and $\Xin$ are identical exactly if the
limiting FD plot is a straight line. They have otherwise no reason to
be related and there is no need to invoke RG calculations 
to ``clarify'' this issue.

\begin{figure}
\begin{center}
\epsfig{file=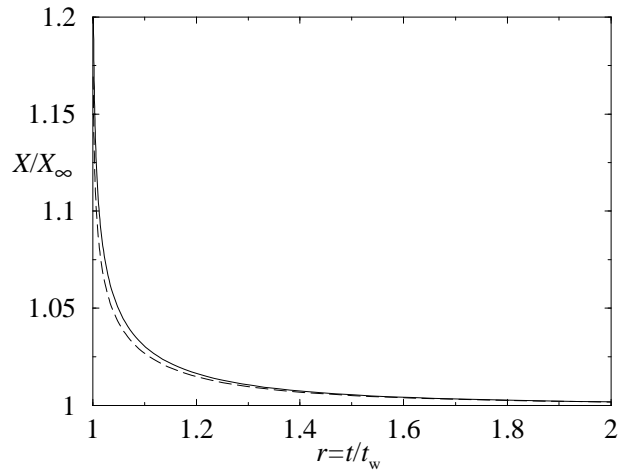,width=8cm}
\end{center}
\caption{RG predictions for $X(t,\tw)/\Xin$ as a function of the
time ratio $\r=t/\tw$, for the two versions of the scaling functions
$F=1+\epsilon^2\Delta F$ (solid line) and $F=\exp(\epsilon^2\Delta F)$
(dashed line). See text for details.
\label{fig:X}
}
\end{figure}

Pleimling's criticism would apply if, somewhat bizarrely, we had
contrived only to collect data in the regime $t\gg\tw$. Such data
would, in a normalized FD plot, fall very close to the plot's end
point at $(1-\Ct,\chit)=(1,Y)$. In an unnormalized plot, the
$t$-dependent stretching of the plot by $C(t,t)$ would then indeed
mean that the data trivially fall on a straight line.  This line would
be $\tw$-independent and have slope $Y$ rather than $\Xin$. To
check for such trivial behaviour, it is sufficient to normalize the
data as explained above.  We re-emphasize that, as
Fig.~\ref{fig:norm_data} shows, our data are not in the regime where
such trivial behaviour is expected, covering a wide range of values of
$1-\Ct$ and remaining well away from the end point of the plot. Our
observation of a close-to-straight line FD plot is therefore not
explained by scaling arguments, and remains highly nontrivial. 
This is transparent from Pleimling's own data~\cite{Pleimling03}: 
one sees that his FD plots in Fig.\ 3 actually show data for which 
the $\tw$-dependence of
$C(t,t)-C(t,\tw)$ (his Fig.~1) and $\chi(t,\tw)$ (his Fig.~2) is 
still significant.

\begin{figure}
\begin{center}
\epsfig{file=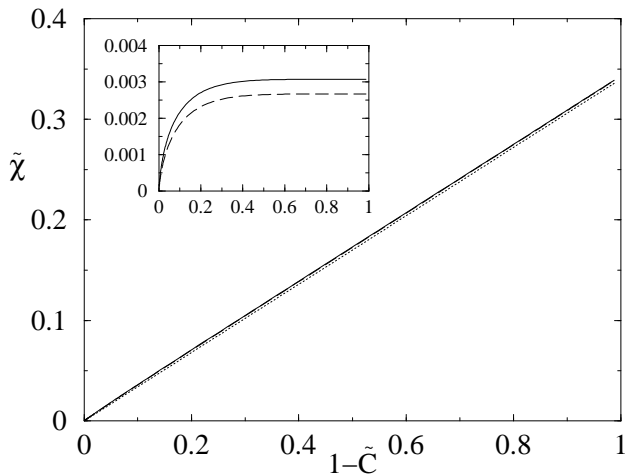,width=8cm}
\end{center}
\caption{RG prediction for long-time limit of the normalized FD plot
for the magnetization. Two versions are shown (solid and dashed),
corresponding to the two choices of scaling function from
Fig.~\protect\ref{fig:X}, but are indistinguishable by eye. Both are
extremely close to a straight line of slope $\Xin$ (dotted). To
make the small differences visible, $\chit-\Xin \Ct$ is shown in the
inset.
\label{fig:RG}
}
\end{figure}

We now comment on Pleimling's statement that our numerical results are
in ``marked contrast'' to the RG calculations
of~\cite{CalGam02}. These calculations give $X(t,\tw)$ as a function
of the time ratio $\r=t/\tw$ in the form
\begin{equation}
X(r) = \Xin \frac{F_R(\r)}{F_{\partial C}(\r)}
\end{equation}
where $F_R$ and $F_{\partial C}$ are appropriate scaling functions for
$R(t,\tw)$ and $(\partial C/\partial\tw)(t,\tw)$, consistent with the
definition~(\ref{equ:FDT}). Both scaling functions are of the form
$F(\r) = 1+\epsilon^2 \Delta F(\r)$, within a second-order expansion
in $\epsilon=4-d$. The extrapolation to $d=2$, $\epsilon=2$ therefore
has a certain arbitrariness.  To ${\mathcal O}(\epsilon^2)$ in the RG
calculation one could replace $F(\r)$ by e.g.\ $\exp(\epsilon^2\Delta
F(\r))$. We show both versions of the resulting RG predictions for
$X(\r)$ in Fig.~\ref{fig:X}. It is clear that $X(\r)$ is close to
$\Xin$ except for $\r\approx 1$; where it does deviate, the RG
predictions also become less reliable. The near-constancy of $X(\r)$
already suggests that the FD plot will be almost straight.

To find the limiting normalized FD plot predicted by RG explicitly, we
combined the RG result for the scaling function $F_{\partial C}(\r)$
with the scaling exponents as quoted by Pleimling to obtain $\upd
\Ct/\upd r$ and then used~(\ref{equ:FDTintr}). The result is shown in
Fig.~\ref{fig:RG} and demonstrates that the RG calculations predict a
limiting FD plot which is extremely close to a straight line. 
Quantitatively, the plot is
shifted upwards from a straight line of slope $\Xin$ by no more
than $0.01\Xin$; its axis ratio $Y$ therefore also lies no more
than $1\%$ above $\Xin$.
Contrary to Pleimling's remark, our numerical data 
are therefore entirely consistent with RG calculations. 

As a final point, we comment on our observation
in~\cite{MayBerGarSol03} that the values of $\Xin$ are, to within
numerical accuracy, identical for the magnetization (a spin
observable) and the energy (a bond observable). While this may be
surprising from the point of view of non-equilibrium 
critical dynamics
\footnote{RG calculations of $\Xin$ for the energy are in progress
for the $O(n)$ model in order to clarify this point (P. Calabrese,
personal communication).}, it is natural if one thinks of
$\Teff=T_c/\Xin$ as an effective temperature which should govern
the long-time non-equilibrium critical dynamics of all (or at least a
broad range of) observables. This is supported by an analysis of the
spherical model, where the values of $\Xin$ for spin and bond
observables do indeed coincide. We will report on a more detailed
investigation of this point in a future publication.

In summary, Pleimling's criticisms of our method of measuring $\Xin$ using
coherent observables do not apply. His reasoning only addresses the
limit $t\gg\tw$, where the normalized correlation function
$\Ct(t,\tw)=C(t,\tw)/C(t,t)$ is negligibly small, while our data are
taken in a regime where $\Ct(t,\tw)$ is of order unity. This is most
easily demonstrated using a normalized FD plot of
$\chi(t,\tw)/C(t,t)$ versus $1-\Ct(t,\tw)$. Our observation that, for
the magnetization in the $d=2$ Ising model quenched to criticality, the
normalized FD plot is close to a straight line therefore remains
nontrivial, and is consistent with RG predictions.

There are two key conclusions of our original
study~\cite{MayBerGarSol03} which we have emphasized throughout this
reply.  First, FD plots for coherent observables are able to reveal
nontrivial two-time dependences in non-equilibrium dynamics, and do so
unambiguously when normalized. Second, FD plots for coherent
observables typically have a wide range where their slope is close to
the asymptotic value $\Xin$.  For measurements of $\Xin$ this
makes them preferable to the traditionally used incoherent
observables, where this range shrinks to zero for long times.

We are grateful to P.\ Calabrese for discussions.  We acknowledge
financial support from CNRS France, EPSRC Grants No.\ 00800822,
GR/R83712/01 and GR/S54074/01, E.U.\ Marie Curie Grant No.\
HPMF-CT-2002-01927, the Glasstone Fund, Nuffield Grant No.\
NAL/00361/G, \"{O}sterreichische Akademie der Wissenschaften, and
Worcester College Oxford.  Some of the numerical results were obtained
on Oswell at the Oxford Supercomputing Center, Oxford University.


\end{document}